\documentclass[11pt,prl,reprint,superscriptaddress]{revtex4-2}

\usepackage{amsmath,amssymb,amsfonts, braket}
\usepackage{graphicx}
\usepackage{bm}
\usepackage[hidelinks]{hyperref}
\usepackage[usenames]{color}
\usepackage[T1]{fontenc}
\usepackage{multirow}
\usepackage{hhline}
\usepackage{booktabs}
\usepackage{makecell}

\usepackage[margin=0.75in]{geometry}

\newcommand{\er}{Er$^{3+}$ }
\newcommand{\ercawo}{Er$^{3+}$:CaWO$_4$ }
\newcommand{\cawo}{CaWO$_4$ }
\newcommand{\eryso}{Er$^{3+}$:Y$_2$SiO$_5 $ }

\newcommand{\upe}{\uparrow_\text{e}}

\begin{document}
\title{Indistinguishable telecom band photons from a single erbium ion in the solid state}

\author{Salim Ourari}
\thanks{These authors contributed equally to this work.}
\affiliation{Department of Electrical and Computer Engineering, Princeton University, Princeton, NJ 08544, USA\looseness=-1}
\author{\L{}ukasz Dusanowski}
\thanks{These authors contributed equally to this work.}
\affiliation{Department of Electrical and Computer Engineering, Princeton University, Princeton, NJ 08544, USA\looseness=-1}
\author{Sebastian P. Horvath}
\thanks{These authors contributed equally to this work.}
\affiliation{Department of Electrical and Computer Engineering, Princeton University, Princeton, NJ 08544, USA\looseness=-1}
\author{Mehmet T. Uysal}
\thanks{These authors contributed equally to this work.}
\affiliation{Department of Electrical and Computer Engineering, Princeton University, Princeton, NJ 08544, USA\looseness=-1}
\author{\\Christopher M. Phenicie}
\affiliation{Department of Electrical and Computer Engineering, Princeton University, Princeton, NJ 08544, USA\looseness=-1}
\author{Paul Stevenson}
\thanks{Present address: Department of Physics, Northeastern University, Boston, Massachusetts 02115, USA}
\affiliation{Department of Electrical and Computer Engineering, Princeton University, Princeton, NJ 08544, USA\looseness=-1}
\author{Mouktik Raha}
\affiliation{Department of Electrical and Computer Engineering, Princeton University, Princeton, NJ 08544, USA\looseness=-1}
\author{Songtao Chen}
\thanks{Present address: Department of Electrical and Computer Engineering, Rice University, Houston, Texas 77005, USA}
\affiliation{Department of Electrical and Computer Engineering, Princeton University, Princeton, NJ 08544, USA\looseness=-1}
\author{\\Robert J. Cava}
\affiliation{Department of Chemistry, Princeton University, Princeton, NJ 08544, USA\looseness=-1}
\author{Nathalie P. de Leon}
\affiliation{Department of Electrical and Computer Engineering, Princeton University, Princeton, NJ 08544, USA\looseness=-1}
\author{Jeff D. Thompson}
\email{jdthompson@princeton.edu}
\affiliation{Department of Electrical and Computer Engineering, Princeton University, Princeton, NJ 08544, USA\looseness=-1}

\begin{abstract}
Atomic defects in the solid state are a key component of quantum repeater networks for long-distance quantum communication \cite{Awschalom2018}. Recently, there has been significant interest in rare earth ions \cite{simon2010,Zhong2018,Kindem2020}, in particular Er$^{3+}$ for its telecom-band optical transition \cite{Dibos2018, ulanowski_spectral_2022, Yang2022}, but their application has been hampered by optical spectral diffusion precluding indistinguishable single photon generation. In this work we implant \er into CaWO$_4$, a material that combines a non-polar site symmetry, low decoherence from nuclear spins \cite{LeDantec2021}, and is free of background rare earth ions, to realize significantly reduced optical spectral diffusion. For shallow implanted ions coupled to nanophotonic cavities with large Purcell factor, we observe single-scan optical linewidths of 150~kHz and long-term spectral diffusion of 63~kHz, both close to the Purcell-enhanced radiative linewidth of 21 kHz. This enables the observation of Hong-Ou-Mandel interference~\cite{Hong1987} between successively emitted photons with high visibility, measured after a 36 km delay line. We also observe spin relaxation times $T_1$ = 3.7 s and $T_2$ > 200 $\mu$s, with the latter limited by paramagnetic impurities in the crystal instead of nuclear spins. This represents a significant step towards the construction of telecom-band quantum repeater networks with single Er$^{3+}$ ions.

\end{abstract}

\maketitle

\onecolumngrid

Long-distance quantum networks are an enabling technology for quantum communication, distributed quantum computing and entanglement-enhanced sensing and metrology \cite{Awschalom2021}. The rate of direct entanglement transmission with photons decreases exponentially with distance, but this can be overcome using quantum repeaters with memories \cite{Briegel1998}. In particular, single atom-like defects in the solid state \cite{Awschalom2018} have been used to demonstrate key milestones including spin-photon entanglement \cite{Togan2010, Greve2012} and single-photon transistors \cite{Sun2018}, remote entanglement of spins \cite{Bernien2013}, entanglement purification \cite{Kalb2017} and memory-enhanced quantum communication \cite{Bhaskar2020}. A challenge to deploying these techniques in long-distance networks is that atomic systems typically operate at transition frequencies outside of the low-loss window of optical fibers, requiring wavelength conversion for long-distance propagation \cite{Li2016, Stolk2022}.

The rare earth ion \er has a telecom-band optical transition at a wavelength of 1.5 $\mu$m that is widely exploited for solid-state optical amplifiers, and in dilute ensembles, as a quantum memory for light \cite{Saglamyurek2015, Craiciu2019}. \er ions can have long spin \cite{Rancic2018, LeDantec2021} and optical \cite{Bottger2009} coherence in a variety of host crystals, a property shared with other rare earth ions \cite{Zhong2015, Ortu2018, Kindem2018}. In recent years, micro- and nano-scale optical resonators have enabled the observation of enhanced single photon emission from \er and other rare earth ions \cite{Dibos2018,Zhong2018,Kindem2020, Ulanowski2022, Yang2022}, which has subsequently enabled single-shot spin readout \cite{Raha2020,Kindem2020} and coupling to nearby nuclear spins that could serve as ancilla qubits \cite{Kornher2020,ruskuc2022,uysal2022}. However, a central challenge to the development of quantum repeaters with single rare earth ions is spectral diffusion, which is particularly pronounced in nanophotonic devices used to achieve fast optical emission from single rare earth ions \cite{Dibos2018,Kindem2020,Yang2022}. To date, indistinguishable single photon emission from a single rare earth ion has not been observed.

Rare earth ions 
 also provide a unique opportunity for materials engineering, as they can be incorporated into a wide range of host crystals while preserving their basic properties, including the optical transition wavelength and spin configuration \cite{thiel2011, Zhong2019, Phenicie2019, Stevenson2022}. An ideal host material would incorporate \er on a non-polar site to suppress linear electric field shifts of the optical transition, and have a low concentration of nuclear spins, other magnetic impurities and particularly trace rare earth ions to allow long spin coherence and low fluorescence background \cite{Ferrenti2020}.

In this work, we demonstrate indistinguishable single photon emission from a single \er ion coupled to a nanophotonic optical cavity. This is enabled by shallow ion implantation of \er into CaWO$_4$, a host material satisfying the above criteria and for which long electron spin coherence has recently been demonstrated in \er ensembles at millikelvin temperatures \cite{LeDantec2021}. By coupling the ions to silicon nanophotonic circuits, we observe individual ions with single-scan optical linewidths of 150 kHz, and emission rate enhancement by a factor of $P=850$ via the Purcell effect. Using a 36 km delay line, we observe Hong-Ou-Mandel (HOM) interference between successively emitted photons with a visibility of $V=80(4)\%$. We also demonstrate spin initialization and single-shot readout with a fidelity $F = 0.972$, as well as the preservation of electron spin coherence for more than 200 $\mu$s, limited by paramagnetic impurities in the sample. This demonstration is a key step for the development of quantum repeaters based on single rare earth ions, and \er in particular.

\begin{figure*}[ht]
	\centering
    \includegraphics[width= \textwidth]{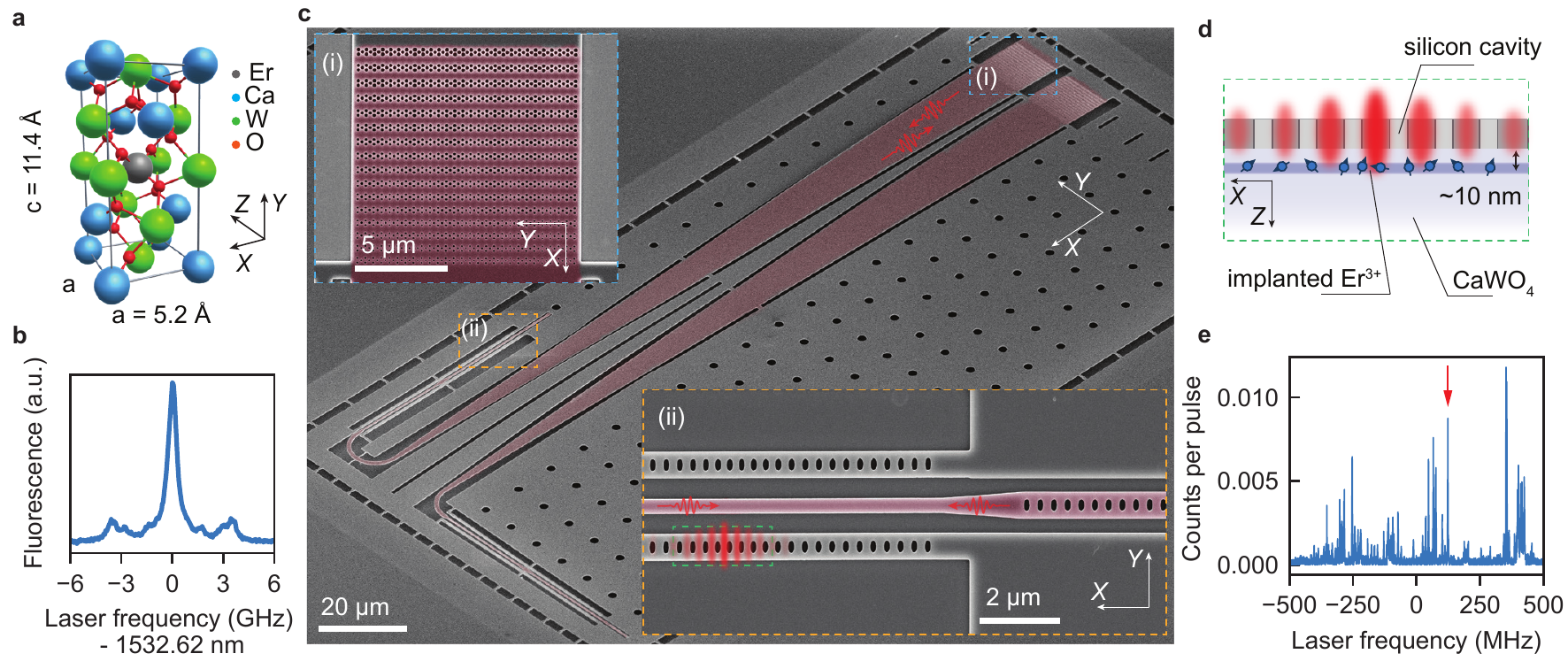}
    \caption{\textbf{Er$^{3+}$:CaWO$_4$ device architecture.} \textbf{a} CaWO$_4$ crystal structure, with a substitutional Er$^{3+}$ impurity in an S$_4$ Ca$^{2+}$ site. \textbf{b} A dense implanted \ercawo ensemble has an inhomogeneous optical linewidth of 730 MHz on the Z$_1$-Y$_1$ transition. In addition to the central peak, we observe hyperfine structure from $^{167}$Er with nuclear spin $I=7/2$. \textbf{c} Scanning electron microscope image of a representative silicon nanophotonic device, consisting of a photonic crystal grating coupler [inset (\emph{i})] that tapers adiabatically into a bus waveguide connected to a photonic crystal nanobeam cavity [inset (\emph{ii})]. \textbf{d}~Erbium ions are implanted targeting a depth of 10 nm, and couple evanescently to the silicon photonic crystal on the surface. \textbf{e} PLE spectrum of \er ions coupled to the cavity, with resolved single ion lines. The red arrow indicates the ion used for subsequent experiments.}
    \label{fig:Fig1}
\end{figure*}

Our samples are produced by introducing erbium into commercially available high purity \cawo using ion implantation with an energy of 35 keV, targeting a depth of 10 nm. In a test sample implanted with a high \er fluence of $1 \times 10^{12}$ ions/cm$^2$, we observe an ensemble optical spectrum at $T=4$ K consistent with substitutional \er on the Ca$^{2+}$ site with S$_4$ symmetry (Fig.~\ref{fig:Fig1}a) \cite{Nassau1963, Bernal1971}. After annealing at $300$ $^\circ$C in air, the inhomogeneous optical linewidth of the Z$_1$-Y$_1$ transition at 1532.63 nm is 730 MHz (Fig.~\ref{fig:Fig1}b). This is comparable to previously reported linewidths in bulk-doped samples (approximately 0.5-1 GHz~\cite{Sun2002, Stevenson2022}), suggesting that the implantation damage is effectively removed by annealing.

To resolve individual ions, we implant a second sample at a lower fluence of $5 \times 10^{9}$ ions/cm$^2$. Single ions are probed using a silicon photonic crystal cavity that is fabricated on a separate silicon-on-insulator wafer, and then bonded to the top surface of the \cawo substrate (Fig. 1c-d) \cite{Dibos2018}. The device and sample are cooled to $T=0.47$~K in a $^3$He cryostat, with optical and microwave access provided with a scanning probe head \cite{Chen2021a}. We probe single ions in the device using photoluminescence excitation (PLE) spectroscopy, by sweeping the frequency of a pulsed laser and observing the time-delayed fluorescence through the cavity with a superconducting nanowire single photon detector (SNSPD). The spectrum contains clearly resolved lines from individual \er ions (Fig. 1e). The number of lines is roughly consistent with the expected number of ions in the cavity area $A=1.3\,\mu$m$^2$, suggesting a high conversion efficiency. The following experiments are performed on the ion indicated by the arrow.

Coupling the \er ion to the cavity allows for optical preparation and measurement of the electron spin. We apply a magnetic field of $|B| = 600$ G to lift the degeneracy of the $S=1/2$ ground and excited states, resulting in two spin-conserving transitions ($A$,$B$) and two spin-flip transitions ($C$,$D$) as shown in Fig.~2a-b (the magnetic moments for the ground and excited state are described in the supplementary information \cite{SI}). Tuning the cavity to the $A$ transition enhances the decay rate of the excited state, shortening the lifetime from 6.3 ms to $\tau = 7.4\,\mu$s, corresponding to a Purcell factor of $P=850$ (Fig.~2c). To enable spin readout, we engineer a cycling transition by selectively enhancing the $A$ transition relative to $D$ by a combination of detuning from the cavity and preferential orientation of the transition dipole moment with respect to the cavity polarization \cite{Raha2020}, resulting in $\Gamma_A/\Gamma_D \approx 1030(10)$~\cite{SI}. Spin initialization is performed by optical pumping on the $A$ or $B$ transitions while simultaneously driving the excited state MW$_e$ transition \cite{Chen2020}. In Fig. 2d, we demonstrate spin initialization and readout with an average fidelity of $F = 0.972$ (Fig. 2d). The combination of high collection efficiency and low background from other \er ions allows for high-contrast optical Rabi oscillations (Fig. 2e). After a $\pi$ pulse, a single photon is detected with a probability $P_1=0.035$, on top of a background count rate of $P_b = P_1/117$. Both $P_1$ and the signal-to-background ratio are larger than what is obtained with frequency converted NV centers by more than an order of magnitude \cite{Stolk2022}, enabled by the high quantum efficiency and collection efficiency of the Er-cavity system.

The linewidth of the spin-conserving transitions is determined using PLE spectroscopy. To avoid optical pumping, the excitation laser has two tones separated by approximately 1 GHz to drive the $A$ and $B$ transitions simultaneously. The typical linewidth of a single scan (1 minute) is approximately 150 kHz, while the line center has an r.m.s. fluctuation of 63~kHz over 12 hours (Fig.~2f). This represents a 100-fold improvement over previously reported linewidths for individual \er ions in nanophotonic cavities \cite{Dibos2018,Chen2020,Yang2022}, and is to our knowledge the narrowest optical transition observed for a solid-state defect in a nanophotonic device. We note that similar linewidths have been observed for single \er ions in 19 $\mu$m thick Y$_2$SiO$_5$ membranes \cite{ulanowski_spectral_2022}. The single-scan linewidth is 7 times larger than the Purcell-enhanced radiative linewidth of the $A$ transition, $\Gamma_r = 1/\tau = 2\pi \times 21.4$ kHz, however, photon echo experiments suggest that this linewidth is dominated by slow dynamics \cite{SI}, such that indistinguishable photon emission may be possible on short timescales or with active feedback.

We perform HOM two-photon interference measurements \cite{Hong1987} on time-delayed photons using an unbalanced Mach-Zehnder interferometer (MZI) with a $\Delta L = 36$ km delay line in one arm (Fig. 3a). By tuning the repetition rate of the excitation pulses to match the delay time of the long arm ($\Delta t = 175\,\mu$s), successive photons may arrive at the final beamsplitter simultaneously, and HOM interference will suppress the probability of detecting one photon at each output if the photons are indistinguishable. Experimentally, we observe strongly suppressed coincidences (Fig.~3b), indicating a high degree of indistinguishability. In a control experiment, we artificially broaden the photon in the short arm using a fiber stretcher driven by a noise source, restoring the coincidence rate expected for distinguishable photons (Fig. 3c). We measure an HOM coincidence rate of $R = 2$~min$^{-1}$, defined as the rate of simultaneous photon detection in the distinguishable photon case, corresponding to a per-shot coincidence probability of $P_c = 8.5\times10^{-6}$.

The indistinguishability is quantified by the visibility $V$~\cite{Santori2002}, given by $V=1- {2A_0}/{ \overline{A}_{|i|\geq 2}}$, where $A_0$ is the integrated counts under the central peak and $\overline{A}_{|i|\geq 2}$ is the average integrated counts in each side peak (Fig. 3b). The visibility is maximized for a coincidence window approaching zero, however the number of photons within this window (the acceptance fraction) will also be small (Fig. 3e). For coincidences with photon detection times $t_1,t_2$ separated by $|t_2-t_1| < 2 \tau$ (corresponding to an acceptance fraction of $63\%$), the raw visibility is over 70\%, rising to 90\% when the accidental coincidences from dark counts and ambient background are subtracted. Integrating under the entire peak in Fig. 3d and subtracting accidental coincidences gives $V=80(4)\%$. The residual distinguishability has a significant contribution (4\%) due to the MZI output beamsplitter ratio deviating from 50:50. Therefore, we conclude that the effective linewidth over hundreds of microseconds is only slightly larger than the radiative linewidth~\cite{SI}.

Lastly, we study the properties of the \er spin, which has the potential to serve as a quantum memory for spin-photon entanglement. In bulk \ercawo, the magnetic moment is anisotropic with $g_c = 1.25$ and $g_a = 8.38$~\cite{Bernal1971}. However, for the individual ions studied in this work, we observe significantly distorted magnetic moments, including a variation of $g$ in the $aa$-plane. These deviations can be reproduced with the inclusion of a small axial crystal field term~\cite{SI}, which may arise from proximity to the surface or the presence of a nearby defect.

The spin relaxation time is $T_1 = 3.7$ s, in line with previous reports \cite{LeDantec2021}, and is limited by the direct process with a $T_1 \propto 1/B^5$ dependence~\cite{SI}. Ramsey and Hahn echo experiments give $T_2^* = 247$ ns and $T_2 = 44\,\mu$s, respectively (Fig.~\ref{fig:Fig4}c-d). An XY$^{64}$ dynamical decoupling sequence allows coherence to be preserved for longer than 200 $\mu$s (Fig.~4e), while also showing collapses and revivals due to the $^{183}$W nuclear spin bath.

The Hahn echo $T_2$ is improved by one order of magnitude from \eryso under similar conditions \cite{Chen2020}, but the coherence is still significantly shorter than predictions based on CCE simulations accounting for the $^{183}$W nuclear spin bath ($I=1/2$, 14.3\% abundance). This implicates paramagnetic impurities in the host crystal or on the surface as the primary source of decoherence, with an inferred density of approximately $3 \times 10^{16}$ cm$^{-3}$~\cite{SI}. Indeed, longer spin echo coherence times of $T_2 = 23$ ms were observed for bulk \er ensembles in CaWO$_4$ in Ref \cite{LeDantec2021} by operating at dilution refrigerator temperatures to freeze out paramagnetic impurities. Although the coherence is not limited by the nuclear spin bath, dips in coherence due to a single strongly coupled $^{183}$W spin are observed in Fig.~\ref{fig:Fig4}d.

The results demonstrated in this work will enable spin-photon entanglement and HOM interference between multiple Er$^{3+}$ emitters with postselection using a narrow coincidence window or active tracking of the transition frequencies. In future work, the radiative linewidth can be further increased using cavities with higher $Q$ \cite{Asano2017} or smaller mode volume \cite{Hu2016}. Furthermore, more careful annealing or surface preparation may reduce the spectral diffusion. The flexibility to incorporate \er via ion implantation, instead of during growth, will allow future exploration of \cawo samples produced and refined using diverse techniques. Reducing the impurity concentration may also improve the optical linewidth: scaling the ground state magnetic linewidth $1/T_2^*$ to the optical transition implies a significant magnetic noise contribution of $2\pi \times 46$ kHz.

In this work, we have demonstrated an engineered material, ion-implanted Er$^{3+}$:CaWO$_4$, that enables indistinguishable single photon generation from a single rare earth ion in the telecom band. We attribute the improved performance to the higher \er site symmetry (compared to previous observations of single \er ions \cite{Dibos2018, Ulanowski2022, Yang2022}). Spectral multiplexing of many ions per node \cite{Chen2020}, using quantum eraser techniques to overcome static frequency differences \cite{Zhao2014}, will enable higher repetition rates over long fiber segments, while simultaneously reducing the coherence time requirements \cite{Collins2007}. Additional storage capacity and functionality may be obtained from ancilla nuclear spin registers, as recently demonstrated for several rare-earth ion systems \cite{ruskuc2022,uysal2022}, and ion implantation may allow for the creation of spatially modulated density profiles with strong magnetic ion-ion interactions.

\emph{Acknowledgements:} We acknowledge helpful conversations with Charles Thiel, Philippe Goldner, and Milo\v{s} Ran\v{c}i\'c. This work was primarily supported by the U.S. Department of Energy, Office of Science, National Quantum Information Science Research Centers, Co-design Center for Quantum Advantage (C2QA) under contract number DE-SC0012704. We also acknowledge support from the DOE Early Career award (for modeling of decoherence mechanisms and spin interactions), as well as AFOSR (FA9550-18-1-0334 and YIP FA9550-18-1-0081), the Eric and Wendy Schmidt Transformative Technology Fund, and DARPA DRINQS (D18AC00015) for establishing the materials spectroscopy pipeline and developing integrated nanophotonic devices. We acknowledge the use of Princeton’s Imaging and Analysis Center, which is partially supported by the PCCM, an NSF MRSEC (DMR-1420541), as well as the Princeton Micro-Nano Fabrication Center.

\emph{Note:} While finalizing this manuscript, we became aware of recent reporting the detection of single \er ions in \cawo using magnetic resonance techniques \cite{wang2023}.

\begin{figure*}[ht]
    \includegraphics[width= \textwidth]{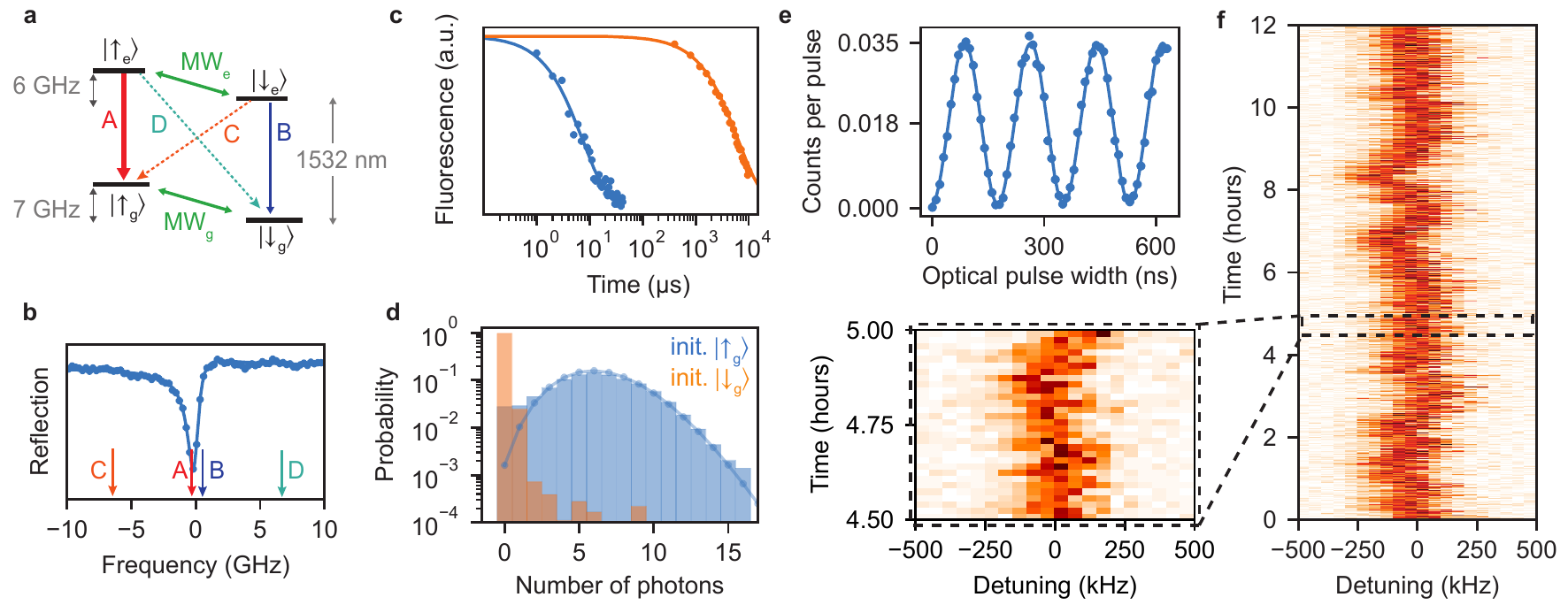} 
    
    \caption{\textbf{Efficient photon collection from a cavity-coupled ion.} \textbf{a} Er$^{3+}$ level structure. In a magnetic field, \er has four distinct optical transitions. The field strength is $|B|=600$~G, oriented in the $aa$-plane, $22^o$ from the $X$-axis.  \textbf{b} Reflection spectrum of the cavity showing a full-width, half-maximum linewidth of $\kappa = 1.0$ GHz ($Q=1.9 \times 10^{5}$), which is tuned into resonance with the $A$ transition. \textbf{c} The lifetime of the $\ket{\upe}$ excited state is reduced to $7.4\,\mu$s (blue), which is 850 times shorter than the bulk lifetime of 6.3 ms (orange). \textbf{d} Histogram of photon counts obtained during spin readout after initializing in $\ket{\uparrow_g}$ and $\ket{\downarrow_g}$. The average readout fidelity is $F=0.972$, using a threshold of one photon. The solid line is a fit to a Poisson distribution with average photon number $\bar{n}= 6.4$.  \textbf{e} Optical Rabi oscillation on transition $A$. The peak single photon emission probability is $P_1 = 0.035$. \textbf{f} Repeated PLE scans show an average single-scan linewidth of 150 kHz, and long-term diffusion of the line center of 63~kHz.}
    
    \label{fig:Fig2}
\end{figure*}

\begin{figure*}[ht]
    \includegraphics[width= \textwidth]{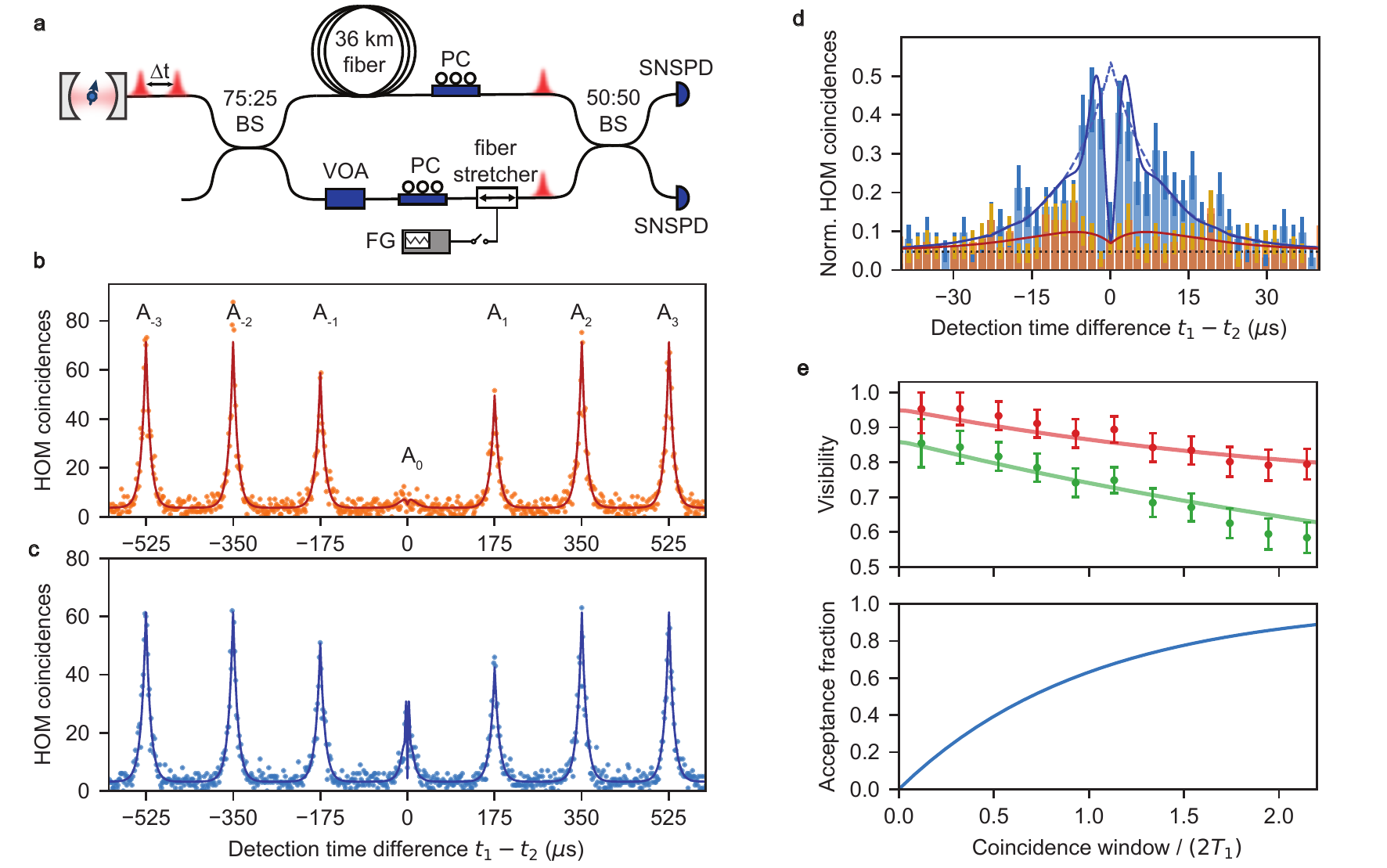} 
    \caption{\textbf{Generation of indistinguishable photons.} \textbf{a} Schematic of the HOM interferometer, indicating beamsplitters (BS), a variable optical attenuator (VOA), polarization controllers (PC) and a fiber stretcher driven by a noise source (FG) to tune the distinguishability. \textbf{b} Histogram of coincidences detected in a 4 hour measurement period. The Hong-Ou-Mandel effect results in a suppressed probability of coincidences at zero delay, indicating indistinguishable single-photon emission. \textbf{c} Histogram of coincidences in a control experiment with a noise source applied to the fiber stretcher, destroying the indistinguishability and Hong–Ou–Mandel interference. \textbf{d} Zoom-in around the zero time delay HOM interference pattern. The red line shows a model including background counts (black dashed line) and pure dephasing of the optical transition. The blue dashed line shows a simple model for the control experiment assuming perfect distinguishability, while the solid blue line shows a model incorporating the finite bandwidth of the noise source~\cite{SI}. \textbf{e} Interference visibility (top) and relative coincidence rate (bottom) as a function of the coincidence window, before (green) and after (red) subtracting the accidental coincidences from the detector and ambient background dark counts.}
    \label{fig:Fig3}
\end{figure*}

\begin{figure*}[ht]
    \includegraphics[]{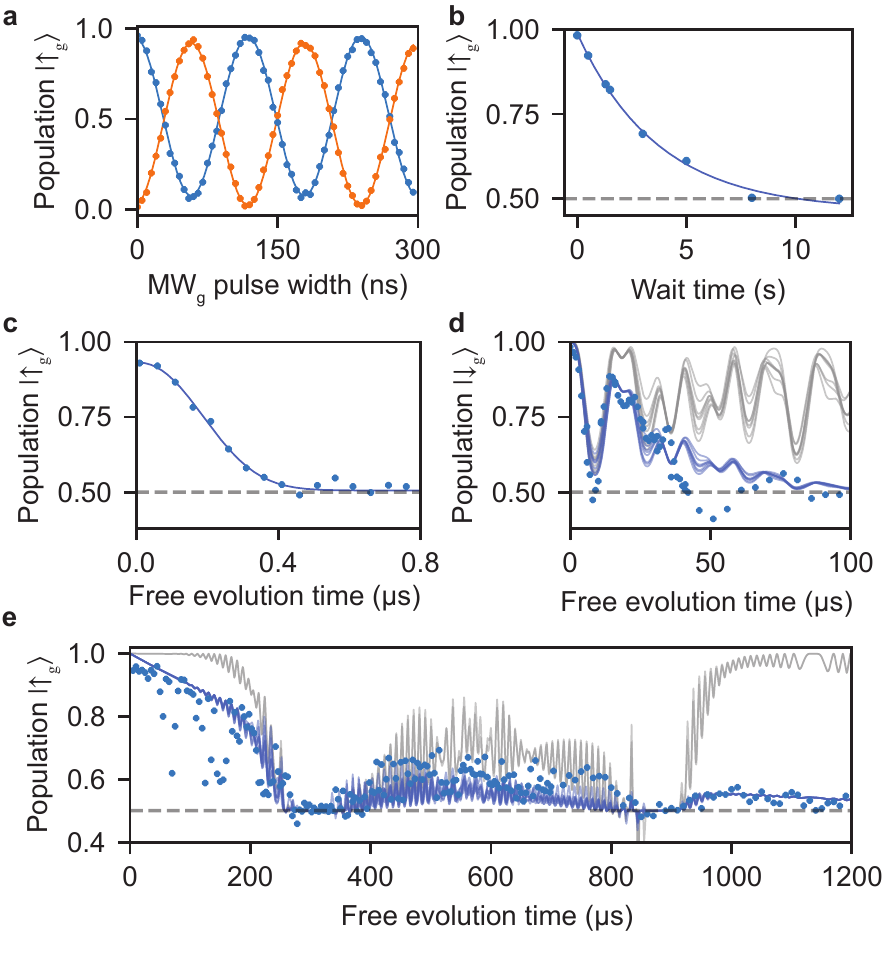} 
    \caption{\textbf{Spin dynamics.}  \textbf{a}~Rabi oscillations after initializing into $\ket{\uparrow_g}$ (blue) or $\ket{\downarrow_g}$ (orange). The spin transition frequency $f_{\text{MW}g}=$~7.0~GHz. \textbf{b}~Spin relaxation after initialization into $\ket{\uparrow_g}$. An exponential fit yields $T_1= 3.7(3)$~s. \textbf{c}~Ramsey measurement. Fitting to $e^{-(t/T_2^*)^n}$ reveals a $T_2^*=247(9)$~ns with $n=2.2(3)$. \textbf{d}~A Hahn echo measurement shows dips in coherence resulting from the $^{183}$W nuclear spin bath. The grey lines show CCE simulations for randomly selected $^{183}$W configurations, where each configuration includes a single strongly coupled $^{183}$W spin required to reproduce the dips. The blue lines include an additional, phenomenological stretched decay with $T_2 = 44\,\mu$s. \textbf{e}~Applying an XY$^{64}$ dynamical decoupling sequence extends the spin coherence to longer times. Here, the grey lines show CCE simulations for the same $^{183}$W bath configurations in panel (d), while the blue lines have an additional phenomenological decay of $460~\mu$s.}
    \label{fig:Fig4}
\end{figure*}

\clearpage
\bibliography{library.bib}

\end{document}


\title{Supplementary information for
 indistinguishable telecom band photons from a single erbium ion in the solid state}
 
\author{Salim Ourari}
\thanks{These authors contributed equally to this work.}
\affiliation{Department of Electrical and Computer Engineering, Princeton University, Princeton, NJ 08544, USA\looseness=-1}
\author{\L{}ukasz Dusanowski}
\thanks{These authors contributed equally to this work.}
\affiliation{Department of Electrical and Computer Engineering, Princeton University, Princeton, NJ 08544, USA\looseness=-1}
\author{Sebastian P. Horvath}
\thanks{These authors contributed equally to this work.}
\affiliation{Department of Electrical and Computer Engineering, Princeton University, Princeton, NJ 08544, USA\looseness=-1}
\author{Mehmet T. Uysal}
\thanks{These authors contributed equally to this work.}
\affiliation{Department of Electrical and Computer Engineering, Princeton University, Princeton, NJ 08544, USA\looseness=-1}
\author{Christopher M. Phenicie}
\affiliation{Department of Electrical and Computer Engineering, Princeton University, Princeton, NJ 08544, USA\looseness=-1}
\author{Paul Stevenson}
\affiliation{Department of Electrical and Computer Engineering, Princeton University, Princeton, NJ 08544, USA\looseness=-1}
\author{Mouktik Raha}
\affiliation{Department of Electrical and Computer Engineering, Princeton University, Princeton, NJ 08544, USA\looseness=-1}
\author{Songtao Chen}
\affiliation{Department of Electrical and Computer Engineering, Princeton University, Princeton, NJ 08544, USA\looseness=-1}
\author{Robert J. Cava}
\affiliation{Department of Chemistry, Princeton University, Princeton, NJ 08544, USA\looseness=-1}
\author{Nathalie P. de Leon}
\affiliation{Department of Electrical and Computer Engineering, Princeton University, Princeton, NJ 08544, USA\looseness=-1}
\author{Jeff D. Thompson}
\email{jdthompson@princeton.edu}
\affiliation{Department of Electrical and Computer Engineering, Princeton University, Princeton, NJ 08544, USA\looseness=-1}

\maketitle
\onecolumngrid

\tableofcontents

\section{Photon collection efficiency}

This section details the efficiency of components in our photonic circuit that impact the total photon collection efficiency. We group these losses into four terms: internal losses in the cavity, in the grating coupler and waveguide taper, transmission through passive optical components, and the quantum efficiency of the SNSPDs. 

The internal losses in the cavity are determined from the reflection spectrum. In particular, the contrast of the reflection on and off cavity resonance $C=(R_{\text{off}}-R_\text{on})/R_\text{off}$ has the form~\cite{Dibos2018}:
\begin{equation}
  C = (1-2 \eta_\text{cav})^2. 
\end{equation} 
The photon extraction efficiency from the cavity $\eta_\text{cav}=\kappa_\text{wg}/\kappa_\text{tot}$ is the ratio of cavity outcoupling rate $\kappa_\text{wg}$ to the total loss $\kappa_\text{tot}=\kappa_\text{wg}+\kappa_\text{int}$, including internal cavity losses $\kappa_\text{int}$. For the device used in this work, we find $\eta_\text{cav} = 0.26$. The grating coupler and waveguide efficiency is estimated by measuring the round-trip optical losses away from the cavity resonance. We compute the efficiency as $\eta_\text{GC}=\sqrt{P_\text{out}/P_\text{in}}$, and find  $\eta_\text{GC} = 0.36$ for the device used in this work. The transmission through passive optical components (beamsplitters, splices, etc.) is measured independently to be $\eta_\text{net} = 0.61$ (the HOM interferometer has additional losses, discussed below). The detection efficiency of the SNSPD is $\eta_\text{det}=0.85$. This gives a combined predicted photon detection probability of $P_1 = \eta_\text{cav} \times \eta_\text{GC} \times \eta_\text{net} \times \eta_\text{det} = 0.049 $. We lose an additional 7\% of the single-ion emission from the finite time required to switch on the SNSPD bias current ($\approx 900$\,ns), which is turned off during the optical excitation pulse to avoid saturating the detector. The predicted photon detection probability is then $P_1 = 0.045$, close to the measured value of 0.035.

In the HOM experiment, the two-photon coincidence probability is decreased by the losses of the MZI delay line. We measured the optical transmission in the $36$~km fiber spool to be $T=0.2$, consistent with an attenuation length $L_a = 21.7$~km. To maximize the coincidence probability and balance the power in the two arms of the interferometer, we use a $75:25$ ratio beamsplitter at the input, such that 0.75 of the power is directed into the long MZI arm, while 0.25 is directed into the short one. In the shorter arm, we utilize a variable optical attenuator set to match the powers before the final MZI beamsplitter input ports. Accounting for both the losses in the fiber spool in the long interferometer arm (matched with a VOA on the short arm), as well as the beamsplitter ratio, the two-photon coincidence probability at $|t_1-t_2|=0$ is $P_\text{HOM} = 0.5\times(0.75P_1 T)^2 = 0.01P_1^2$. Here, the factor of 0.5 accounts for the requirement that the early (late) photon must pass through the long (short) interferometer arm. For $P_1=0.035$ we get $P_\text{HOM}= 1.4\times 10^{-5}$, close to the measured value of $8.5\times 10^{-6}$.
%


\section{CaWO$_4$ sample preparation}

The CaWO$_4$ substrates used in this study were procured from SurfaceNet GmbH with a single sided epi polish and (100) orientation (\emph{i.e.}, with a surface spanned by $a$ and $c$ axes). According to the vendor, the crystals were grown using 99.9999\% purity precursors and sold as ``high purity CaWO$_4$.'' Polished samples were implanted with erbium (II-VI Inc.) using an energy of 35 keV which corresponds to a target depth of 10 nm (calculated by Stopping-Range of Ions in Matter simulations~\cite{ziegler_srim_2010}). Samples were prepared using two different \er concentrations, a high density sample used for ensemble spectroscopy utilizing a fluence of $1 \times 10^{12}$ ions/cm$^{2}$ and a low density sample used for single ion spectroscopy implanted with a fluence of $5 \times 10^{9}$ ions/cm$^{2}$. Subsequent to implantation, samples were annealed in air at a temperature of $T=300$~$^\circ$C for 1 hour, with a heating rate of $T=300$~$^\circ$C/hour. Annealing healed implantation damage and improved site occupation of the S$_4$ point-group symmetry substitutional site.

Silicon photonic crystal cavities are prepared as described previously \cite{Chen2021a}. The cavities are oriented on the substrate such that the predominant electric field polarization is parallel to the CaWO$_4$ $c$-axis.

\section{Site-selective excitation spectroscopy}

\begin{figure*}[h]
	\centering
    \includegraphics[]{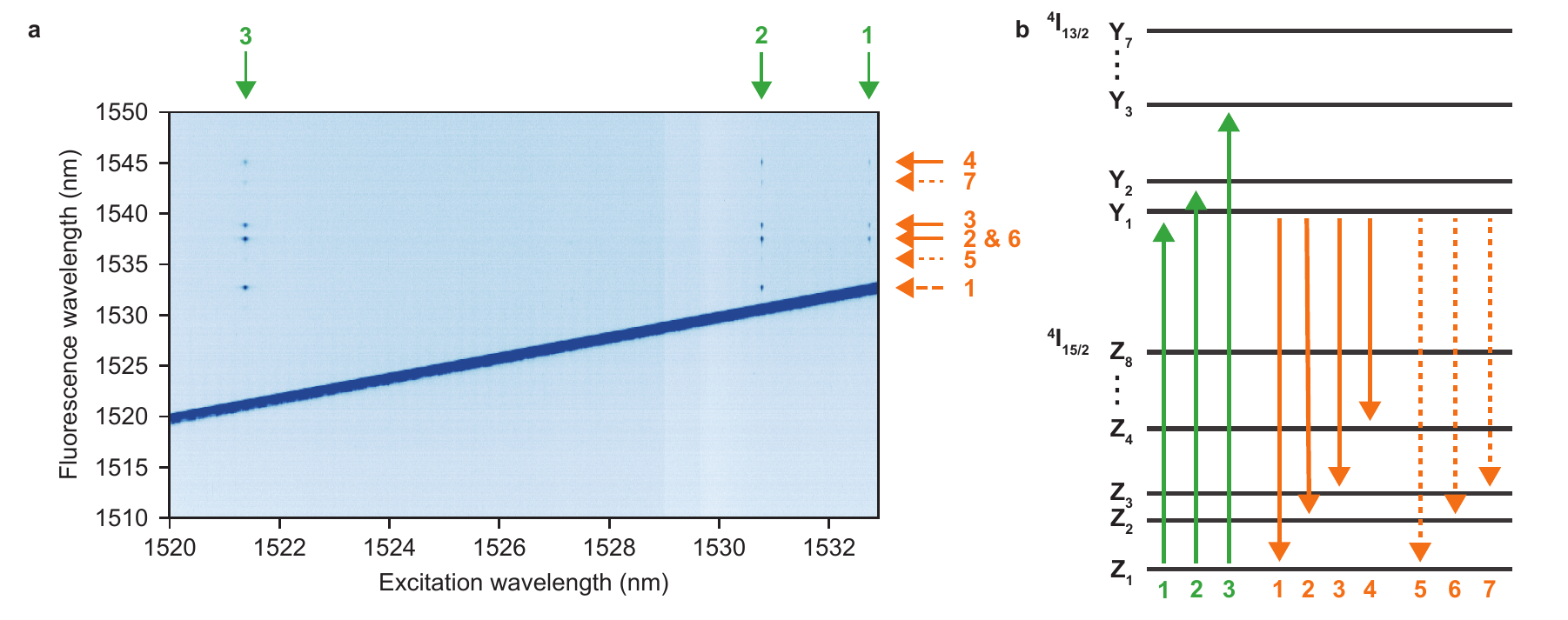} 
    \caption{\textbf{Ensemble spectroscopy of \ercawo}. \textbf{a}~ Site-selective excitation spectrum of \ercawo. Green arrows indicate when the excitation laser is resonant with different excited state crystal-field levels, where the corresponding excitation path is labeled using the matching number in (b). Solid orange arrows denote the fluorescence energies due to decay from the $^4$I$_{13/2}$Y$_1$ level, whereas the dashed arrows denote decay from the $^4$I$_{13/2}$Y$_2$ level. The prominent line with unity gradient corresponds to laser scatter and has been re-scaled in intensity for clarity. \textbf{b}~The inferred energy level structure of \ercawo. In total, the performed spectroscopy yielded energies of four $^4$I$_{15/2}$ and three $^4$I$_{13/2}$ levels.} 
    \label{fig:ex_spect}
\end{figure*}

In order to confirm that implanted \er ions substitute at a site with S$_4$ symmetry we performed a site-selective excitation spectroscopy measurement using the high density sample cooled to 4~K. A tunable narrowband laser was employed in conjunction with an optical chopper to generate a train of excitation pulses. Using a second chopper, fluorescence was collected out of phase with the excitation laser, dispersed using a monochromator, and detected with an InGaAs detector array. By performing a fluorescence measurement while sweeping the excitation laser a map of site-specific excitation and emission frequencies was obtained (Fig.~\ref{fig:ex_spect}a). Table~\ref{tab:energy_levels} summarizes the measured transition energies of four $^4$I$_{15/2}$ and three $^4$I$_{13/2}$ levels. The resulting energy level structure is shown in Fig.~\ref{fig:ex_spect}b, and found to be in close agreement with the transition energies previously reported for bulk doped samples \cite{Bernal1971}, confirming the S$_4$ site assignment. The spectrum does not show any detectable fluorescence at other wavelengths within our scan range, suggesting that this is the only \er incorporation site.
\setlength{\tabcolsep}{20pt}
\begin{table}[ht]
\caption{\label{tab:energy_levels} Transition energies of implanted \ercawo~ determined using site-selective excitation spectroscopy. All values are in cm$^{-1}$. Transitions to Z$_n$ and Y$_n$ for $n$ greater than what is shown were not observed due to limitations in detection sensitivity. The observed transition energies are in close agreement with values obtained for bulk doped samples \cite{Bernal1971}.} 
\begin{tabular}{ ccc }
  \hline
  n & $^4$I$_{15/2}$Z$_n$ & $^4$I$_{13/2}$Y$_n$  \\ 
  \hline
  1 & 0 & 6524.4 \\
  2 & 20.3 & 6532.9 \\
  3 & 25.9 & 6573.2 \\
  4 & 52.0 & * \\
  \hline
\end{tabular}
\end{table}%

We note that compared to the optical excited state lifetime of 6.3~ms the Y$_n$ crystal field levels thermalize rapidly, such that an identical fluorescence spectrum is obtained independent of which $^4$I$_{13/2}$ level is excited (green arrows in Fig.~\ref{fig:ex_spect}). Furthermore, while the fluorescence spectrum is dominated by decay from the $^4$I$_{13/2}$Y$_1$ level, a small fraction of fluorescence originates from the $^4$I$_{13/2}$Y$_2$ level. This leads to two sets of fluorescence patterns with identical energy splittings (but different intensities), overlaid with an offset given by the $^4$I$_{13/2}$Y$_1$-$^4$I$_{13/2}$Y$_2$ splitting (indicated using dashed arrows in Fig.~\ref{fig:ex_spect}). 


\section{Spin state initialization and readout}

To achieve fast and efficient spin state initialization we follow the approach used in Ref.~\cite{Chen2020}. This initialization scheme consists of using optical $\pi$ pulses resonant with either the $A$ or $B$ transition, each followed by a microwave pulse resonant with the excited state spin transition, MW$_e$. For example, to initialize into the $\ket{\downarrow_g}$ state we alternate $\pi$ pulses resonant with the optical $A$ and MW$_e$ transitions. The number of pulse pairs used for the HOM and spin dynamics experiments was 1 and 10, respectively.

For spin state readout, we used a sequence of $n$ optical $\pi$ pulses resonant with the $A$ transition each followed by a fluorescence collection window. By varying the number of readout pulses we found that the spin state readout fidelity is maximized for $n=195$, and applying further pulses would reduce the readout fidelity due to optical pumping. After optimizing the number of readout pulses, a threshold was set for the total number of photons collected after $n$ pulses ($N_{\text{thresh}}$) to discriminate between the $\ket{\downarrow_g}$ and $\ket{\uparrow_g}$ states. We found that a threshold of $N_{\text{thresh}}=1$ gives the highest readout fidelity; that is, if we obtained on average one photon or more after the $n$ pulses then the spin state was assigned to $\ket{\uparrow_g}$, and if we obtained no photon then the spin state was assigned to $\ket{\downarrow_g}$.

\section{Engineering an optical cycling transition}

As noted in Ref.~\cite{Raha2020} for Er$^{3+}$:YSO, the cyclicity of the optical transition depends strongly on the magnetic field orientation since changing the field changes the atomic transition dipole moment with respect to the cavity polarization. For the case where the cavity linewidth is large compared to the Zeeman splitting, the cyclicity is maximized when the spin-flip transitions $C$, $D$ are orthogonal to the cavity polarization. In this work, by combining a larger $C$-$D$ splitting and a narrower optical cavity than in Ref.~\cite{Raha2020} we were able to tune the $A$ transition into resonance with the cavity while simultaneously keeping the spin-flip transitions $C$ and $D$ detuned from the cavity (see Fig. 2b). Consequently we optimize the magnetic field orientation to maximize cyclicity. From this we found the optimal field orientation to be in the $aa$-plane, rotated $22^o$ from the $X$-axis.

\begin{figure*}[h]
	\centering
    \includegraphics[]{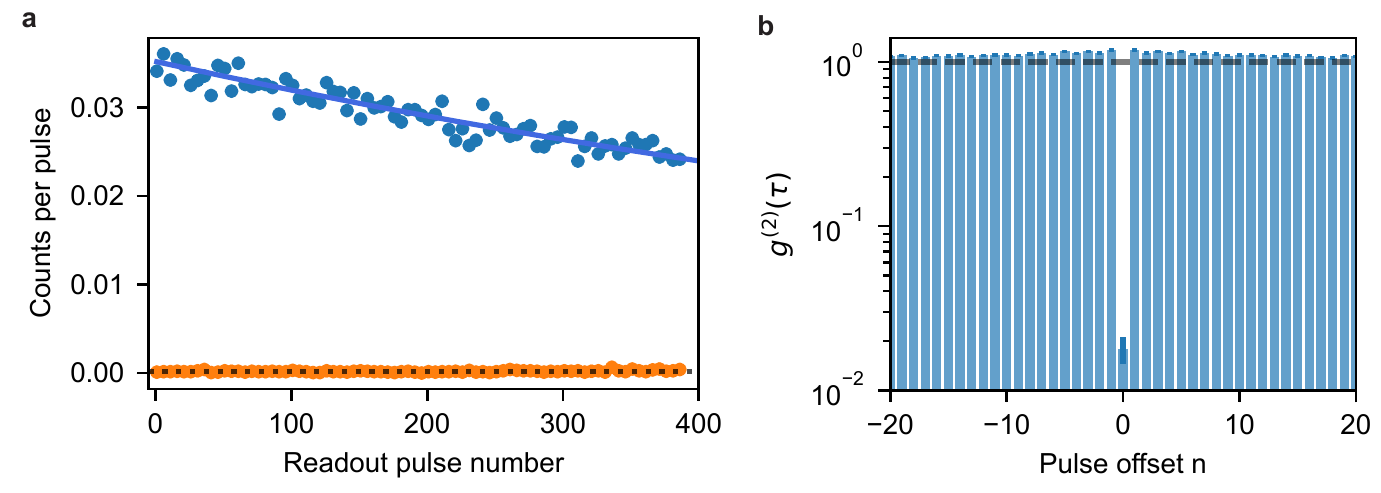} 
    \caption{\textbf{Measuring the cyclicity.} \textbf{a} Decay of the $A$ transition fluorescence count rate from optical pumping after initializing into $\ket{\uparrow_g}$ (blue). The count rate decays as $e^{-n/C}$ revealing a cyclicity of $C=1030(10)$. The orange trace corresponds to the same readout pulse sequence after initializing into $\ket{\downarrow_g}$. \textbf{b} Intensity autocorrelation $g^{(2)}(\tau)$ of the $A$ transition showing strong suppression of the zero-delay peak with $g^{(2)}(0)= 0.018(3)$.}
    \label{fig:cyclicity}
\end{figure*}

To probe the cyclicity, the ion was initialized into $\ket{\uparrow_g}$ and subsequently read out using 400 pulses. Each readout pulse consisted of an optical $\pi$ pulse resonant with the $A$ transition followed by a fluorescence collection window. Due to the finite cyclicity, the fluorescence count rate decays with increasing readout pulse number. This yielded a cyclicity of $C = \Gamma_A/\Gamma_D = 1030(10)$ (Fig.~\ref{fig:cyclicity}a).

To establish the single photon emission we calculate the intensity autocorrelation $g^{(2)}(\tau)$ as shown in Fig.~\ref{fig:cyclicity}b. At zero offset pulse delay, we observe $g^{(2)}(0)=0.018(3)$, showing strong suppression of multi-photon emission events and thus a high purity of single photon emission.

\section{Photon echo investigation of the optical coherence}
\begin{figure*}[h]
	\centering
    \includegraphics[]{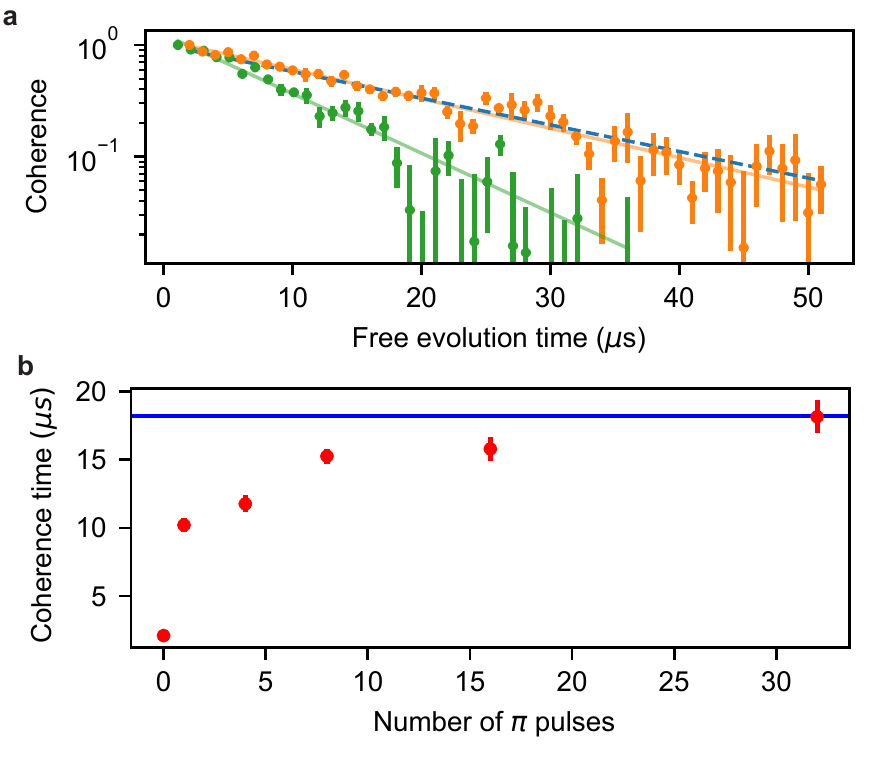} 
    \caption{\textbf{Optical coherence of \ercawo}. \textbf{a} An optical Hahn echo measurement (green) reveals an optical coherence of $T_2=10.2~\mu$s. Applying $XY^{32}$ dynamical decoupling sequence (orange) extends the optical coherence to the radiative limit (blue dashed line) $T_2=18~\mu$s, at a field where the optical $T_1=9.1(3)~\mu$s. \textbf{b} Optical coherence scaling with the number of refocusing pulses of $XY$ dynamical decoupling sequences (red) compared to the lifetime limit (blue). } 
    \label{fig:PhotonEcho}
\end{figure*}

We probe the coherence of the optical $A$ transition using the photon echo technique. After initialization of the spin state, we utilized an excitation sequence consisting of three optical pulses $\pi$/2~-~$\pi$~-~$\pi$/2 separated by a waiting time $\tau$ (for a total free evolution time 2$\tau$), followed by a fluorescence collection window. The refocusing $\pi$ pulse in the middle of the sequence removes the slowly varying inhomogeneous dephasing. For each evolution time, we swept the phase of the last $\pi$/2 pulse and fitted the fluorescence intensity change against the phase angle using a sine function. The fitted oscillation amplitude is proportional to the two-level coherence. This yielded a coherence time of $T_2 = 10.2~\mu$s for a Hahn echo sequence (Fig.~\ref{fig:PhotonEcho}a blue points). To further filter out higher frequency noise, we increased the number of refocusing pulses using an $XY^N$ dynamical decoupling sequence (see Fig.~\ref{fig:PhotonEcho}b) and reached a radiatively limited coherence time of 18~$\mu$s for $N$~=~32 (see Fig.~\ref{fig:PhotonEcho}a orange points). We note that in this experiment the emission lifetime is $T_1=9.1$~$\mu$s, different from $7.4$~$\mu$s recorded using time-resolved PLE shown in the main text. The results of this experiment implicate slow spectral diffusion as the main source of decoherence in our system. In the case of the Hahn echo sequence, the recorded $T_2$ time is approximately a factor of two from the lifetime limit $T_2/(2T_1)=0.56$. 

\section{HOM fit functions}
In this section, we derive the two-photon interference functions used to model the HOM histograms in the main text. In particular, we consider two dephasing mechanisms, which allow us to estimate the possible bounds on the emission linewidth based on the photon visibility determined from the HOM experiment. Finally, we extend the model to simulate the reference HOM measurement, where the photons are made distinguishable by periodically changing the phase of one of the two interfering photons.

To model the HOM zero-delay time peak shape, we follow the derivation of Ref~\cite{Kambs2018}. We assume that single photon spatio-temporal wave functions have an exponential form:
\begin{equation}
\zeta (t) = H(t) \cdot \exp \left( -\frac{t}{2T_1} - i[\omega(t) t + \phi(t)] \right), 
\end{equation}
where $H(t)$ is the Heaviside function, $T_1$ is the radiative lifetime, $\omega(t)$ is the photon frequency and $\phi(t)$ is the phase. In an ideal case, the frequency and phase of the photon are constant over time, so the photon coherence is lifetime limited. A time-dependent frequency and phase noise lead to decoherence. Typically it is assumed that such perturbations yield random walks in phase and frequency at two distinct timescales leading to two different physical descriptions. The first regime is pure dephasing, caused by fast frequency jumps accumulating phase on a timescale shorter than $T_1$. The second regime is described by spectral diffusion, primarily attributed to slow frequency drift on a timescale considerably longer than $T_1$. It is worth noting that this time-scale distinction is somewhat artificial, as it is known that different noise sources have continuous power spectral densities~\cite{kuhlmann_transform-limited_2015}. Still, we perform this analysis to study limiting cases. It can be shown that for single photons passing through an unbalanced MZI with a delay equal to the photon generation rate, the HOM histogram peak areas $A_n$ will be given by: $A_{|i| \geq 2}=A$, $A_{1}=A(1-R^2)$, $A_{-1}=A(1-T^2)$ and $A_{0}=A(R^2+T^2 - 2RTV_\text{int})$~\cite{loredo_scalable_2016}. Here, $A$ is the Poissionian peak coincidence level, $R$/$T$ is the reflection/transmission coefficient of the output MZI beamsplitter, and $V_{int}$ is the emitter visibility. In such cases, the HOM two-photon interference coincidence probability function can be described by
\begin{equation}
\begin{split}
\label{eq:P_HOM}
P(\tau) =& P_\text{dc} + A \sum_{k=2}^N { e^{-\frac{|\tau \pm k t_\text{rep}|}{T_1} } } + A  (1-R^2)e^{-\frac{|\tau + t_\text{rep}|}{T_1}} + A(1-T^2)e^{-\frac{|\tau - t_\text{rep}|}{T_1}} \\ &+ A e^{-\frac{|\tau|}{T_1}} (R^2+T^2 - 2RT\cdot F(\tau) ),
\end{split}
\end{equation}
where $P_\text{dc}$ is the coincidence level related to SNSPD dark counts and ambient light counts, and $t_\text{rep}$ is the pulse repetition period. Further, $F(\tau)$ is an integral defined as~\cite{Kambs2018}
\begin{equation}
F(\tau) = \int_{-\infty}^{\infty} dt_0 \cos \left[ \Delta \omega(\tau,t_0)\tau + \Delta \phi(\tau,t_0) \right],
\end{equation}
where $\Delta \phi(\tau,t_0)$ is a the phase difference and $\Delta \omega(\tau,t_0)$ is the frequency difference between two interfering photons. If frequency and phase fluctuations are assumed to follow Gaussian distribution functions, $F(\tau)$ is given by~\cite{Kambs2018}
\begin{equation}
F(\tau) = \exp \left(-\frac{2|\tau|}{T_\text{dep}} - \sigma^2 \tau^2 \right),
\end{equation}
where $T_\text{dep}$ is the dephasing time $1/T_\text{dep}=1/T_2 -1/(2T_1)$, and $\sigma$ is the inhomogeneous linewidth broadening related to slow spectral diffusion. 

First, we consider the case where fast spectral diffusion dominates (Lorentzian broadening), for which 
\begin{equation}
\label{eq:F_Lor}
F_\text{lor}(\tau) = \exp \left(-\frac{2|\tau|}{T_\text{dep}} \right).
\end{equation}
This allows for the evaluation of the HOM histogram central peak area
\begin{equation}
A_0 = A \int_{-\infty}^{\infty} d\tau  \left( R^2+T^2 - 2RT  e^{-\frac{2|\tau|}{T_\text{dep}}} \right) e^{-\frac{|\tau|}{T_1}} = A\left[  2T_1(R^2 + T^2) - 4RT\frac{T_1 T_\text{dep}}{2T_1 + T_\text{dep}} \right],
\end{equation}
and the off-center peak area $A_{|i| \geq 2 }$  
\begin{equation}
A_{|i| \geq 2 } = A \int_{-\infty}^{\infty} d\tau e^{-\frac{|\tau|}{T_1}} = 2AT_1,
\end{equation}
giving a visibility of 
\begin{equation}
V = 1 - \frac{2A_0}{A_{|i| \geq 2 }} = 1 - 2\left( R^2 + T^2 \right) + 4RT\frac{T_\text{dep}}{2T_1 + T_\text{dep}}=1 - 2\left( R^2 + T^2 \right) + 4RT\frac{T_{2}}{2T_1}.
\end{equation}
Note that this definition of visibility contains information about both the MZI interferometer imperfections (BS $R:T$) and the intrinsic indistinguishability of the emitted photons $V_\text{int}=T_2/(2T_1)$. In our experiment $R:T=0.43:0.57$ is determined from the imbalance between $A_{-1}$ and $A_{1}$ peak areas in the HOM histogram (see Fig. 3b in the main text), contributing to a decrease in visibility of around 4\%. By evaluating the integrated counts $A_0$ and $A_{|n| \geq 2 }$ in the HOM experiment, we obtain a visibility of 80\% after the dark and ambient light counts are subtracted. This allows us to estimate $V_\text{int}=0.84$, and further $T_2=0.84\times2T_1=15.3$~$\mu$s. Using Eqs.~\eqref{eq:P_HOM} and~\eqref{eq:F_Lor} with $T_2=15.3$~$\mu$s we model the HOM histogram in the main text (Fig. 3b,d). Furthermore, using this model, we can estimate the bound on the emission linewidth at the timescale of $100-200$~$\mu$s following
\begin{equation}
    \nu_{L} =  \frac{1}{\pi T_2} = \frac{1}{2\pi T_1 V_\text{int}},
\end{equation}
which yields Lorentzian broadening of $\nu_{L} = 20.6$~kHz. Note that in the case of Fourier limited photons, one obtains $\nu_{LF} = 1/(2\pi T_1)=17.3$~kHz.  

In the case where slow dynamics dominate decoherence (Gaussian broadening), the function $F(\tau)$ is simplified to $ F_\text{gau}(\tau) = \exp \left( - \sigma^2 \tau^2 \right)$, such that
\begin{equation}
    A_0 = A\left[ 2T_1(R^2 + T^2) - 2RT \frac{ \sqrt{\pi} e^{\frac{1}{4\sigma^2T_1^2}} \erfc \left( \frac{1}{2\sigma T_1} \right)}{\sigma} \right]. 
\end{equation}
This leads to a visibility given by
\begin{equation}
    V = 1 - 2\left( R^2 + T^2 \right) + 2RT \frac{ \sqrt{\pi} e^{\frac{1}{4\sigma^2T_1^2}} \erfc \left( \frac{1}{2\sigma T_1} \right)}{\sigma T_1},    
\end{equation}
where the intrinsic emitter visibility is given by
\begin{equation}
\label{eq:VintGuass}
    V_\text{int} = \frac{ \sqrt{\pi} e^{\frac{1}{4\sigma^2T_1^2}} \erfc \left( \frac{1}{2\sigma T_1} \right)}{2\sigma T_1}.    
\end{equation}
Using $V_\text{int}=0.84$ we get $\sigma$ equal to 0.039~MHz, which corresponds to a Gaussian broadening of $\nu_{G} = 2\sqrt{2ln2} \sqrt{2} \sigma=21$~kHz. Note that the estimated $\nu_{G}$ is on the order of the Fourier limit of 17.3~kHz, such that the resultant emission line-shape will be described by a Voigt profile with a total width of
\begin{equation}
\label{eq:Voigt}
    \nu_{V} = \frac{0.535}{2\pi T_1} + \sqrt{ \frac{0.217}{(2\pi T_1)^2} + \nu_{G}^2},
\end{equation}
equal to a linewidth of 31.4~kHz. This bounds the emission linewidth to be in the range $20.6-31.4$~kHz for a timescale of 175.2~$\mu$s.

Next, we consider the case of the reference HOM measurement in Fig. 3c. Typically, the distinguishability in HOM experiments is tuned by rotating the polarization of one of the photons. However, our SNSPDs have a strongly polarization-dependent detection efficiency, which complicates the interpretation of such a measurement. Therefore, we instead make the photons distinguishable by artificial spectral broadening, achieved by rapidly modulating the path length of one of the interferometer arms with a large amplitude. Specifically, the phase of the photons traveling through the shorter MZI arm is modulated using a triangle wave with frequency $\frac{\omega_{m}}{2\pi}=43$~kHz and amplitude $A_m = 0.75\pi$. Consequently, the phase difference can be described directly by
\begin{equation}
\Delta \phi(\tau,t_0) = \frac{2A_m}{\pi} \left[ \arcsin \left( \sin \left( \omega_m (\tau+t_0) \right) \right) - \arcsin \left( \sin \left( \omega_m t_0 \right) \right) \right].
\end{equation}
In this case we can assume that the HOM zero-delay peak area will be dominated by the external phase modulation so that we can omit the $\Delta \omega$ term, and $F(\tau)$ will only depend on $\Delta \phi(t_0,\tau)$
\begin{equation}
F_\text{mod}(\tau) = \int_{-\infty}^{\infty} dt_0 \cos \left(  \frac{2A_m}{\pi} \left[ \arcsin \left( \sin \left(\omega_m (\tau+t_0) \right) \right) - \arcsin \left( \sin \left(\omega_m t_0 \right) \right) \right]  \right).
\end{equation}
The periodic modulation of the phase difference with $\tau$ will translate into a quantum-beat-like signal with a distinct dip at zero detection time difference. To fit the HOM data in Fig. 3c,d of the main text, we evaluate $F_\text{mod}(\tau)$ numerically. The best fit to the experimental data is obtained for a modulation amplitude of 0.73(5)$\pi$ where the modulation frequency was fixed to the value used in the experiment, 43~kHz.

\section{Distortion of the magnetic moment tensor}
\label{sec:g-factor}
The magnetic moments of individual single ions changed appreciably from ion-to-ion as well as from the previously recorded ground state $g$-tensor \cite{Bernal1971}. To investigate this, the magnetic response of two single ions (different from the single ion investigated in the main text) was fully characterized by probing the optical transition frequencies of the four transitions $A$, $B$, $C$, and $D$ for a range of field orientations using a magnetic field magnitude of $50$ G, with the results shown in Tab.~\ref{tab:single-g-vals}. 
\begin{table}[ht]
\caption{Single-ion $g$-tensors measured for two separate ions. Due to a small distortion of the S$_4$ point-group site, the single-ion magnetic moments do not satisfy $g_x = g_y = g_{\perp}$. We note that in this and subsequent sections we use the convention where the $z$ axis points along the crystallographic $c$ axis, which is different to the coordinate system used in the main text.}
\centering
\begin{tabular}{ rcccccc } 
  \hline
  & \multicolumn{3}{c}{Ground state} & \multicolumn{3}{c}{Excited state} \\
   & $g_x$ & $g_y$ & $g_z$ & $g_x$ & $g_y$ & $g_z$\\ 
  \hline
  Ion 1 & 8.5 & 7.6 & 1.7 & 7.3 & 6.9 & 1.8 \\
  Ion 2 & 8.6 & 7.9 & 2.5 & 7.6 & 6.9 & 2.3 \\
  \hline
\end{tabular}
\label{tab:single-g-vals}
\end{table}

In order to gain a deeper understanding of the $g$-tensor anisotropy, an effective crystal-field Hamiltonian was used to model the intra-$4f$ transitions of Er$^{3+}$:CaWO$_4$. The complete Hamiltonian has the form
\begin{equation}
   H = H_{\mathrm{FI}} + H_{\mathrm{CF}} + H_{\mathrm{Z}}, 
   \label{eqn:h_cf}
\end{equation}
where $H_{\mathrm{FI}}$ corresponds to the free-ion components of the Hamiltonian, $H_{\mathrm{CF}}$ accounts for the interaction of the valance electrons with the host material, and $H_{\mathrm{Z}}$ is the Zeeman Hamiltonian. The free-ion Hamiltonian used is \cite{carnall1989}
\begin{equation}
   H_{\mathrm{FI}} = E_{\mathrm{AVG}} + \sum_{1,2,3} F^k f_k + \zeta_{4f} A_{\mathrm{SO}} + \alpha L(L+1) + \beta G(G_2) + \gamma G(R_7) + \sum_{i = 2,3,4,6,7,8} T^i t_i.
   \label{eqn:h_fi}
\end{equation}
Noting only the most significant contributions, $E_{\mathrm{AVG}}$ accounts for the spherically symmetric one-electron component, $F^k$ are the Slater parameters, $f^k$ are the electrostatic repulsion with angular dependence, $\zeta_{4f}$ is the spin-orbit coupling term, and $A_{\mathrm{SO}}$ is the spin-orbit coupling operator. The remaining contributions are higher-order and the reader is referred to Ref.~\cite{wybourne1965} for details. To model the contribution of the S$_4$ point-group symmetry crystal-field the following Hamiltonian was used
\begin{equation}
    H_{\mathrm{CF}} = B^2_0 C^{(2)}_0 + B^4_0 C^{(4)}_0 + B^6_0 C^{(6)}_0 +  B^4_4(C^{(4)}_4 + C^{(4)}_{-4}) + B^6_4 (C^{(6)}_4 + C^{(6)}_{-4}), 
    \label{eqn:cf_h}
\end{equation}
with the coefficients $B^k_q$ the crystal-field parameters and $C^{(k)}_q$ spherical-tensor operators in Wybourne's normalization. We note that the above $B^k_q$ are all real with the exception of $B^6_4$, which is complex. The above Hamiltonian was fitted to data obtained using site-selective fluorescence spectroscopy of the $^4$I$_{13/2} \to ^4$I$_{15/2}$ transitions as well as the barycenter of higher-energy transitions up to $^4$S$_{3/2}$ obtained from literature \cite{Bernal1971}. Furthermore, the fit was optimized to reproduce the ensemble $^4$I$_{15/2}$Z$_1$ and $^4$I$_{13/2}$Y$_1$ $g$-tensors. 

Using the crystal-field parameters obtained via the method outlined above as a starting point, a fit was then performed to the $g$-tensor of both Ion 1 and Ion 2 to determine the crystal-field environment local to each ion. We hypothesize that the perturbation of the S$_4$ point-group symmetry was caused by a single dominant defect, potentially due to the substrate surface or some form of remote charge compensation. Consequently, an axial term $\bar{B}^2_0$ oriented at an arbitrary angle with respect to the crystal-field quantization axis was introduced, and both the magnitude as well as the orientation were varied to reproduce the observed ground and excited state $g$-tensors. This assumed that the crystal-field potential due to the defect is superposable with the nominal crystal-field potential~\cite{newman1971}. 

For Ion 1, the observed $g$-tensor was reproduced by an axial term $\bar{B}^2_0 = 9.3$ cm$^{-1}$, rotated by an Euler rotation from the quantization axis with $\alpha= 90.4^\circ$, $\beta = 265^\circ$ and $\gamma = 0^\circ$, following the Euler angle convention of Messiah~\cite{messiah1961}. Similarly, Ion 2 required an axial term $\bar{B}^2_0 = 10.7$~cm$^{-1}$, rotated by an Euler rotation from the quantization axis with $\alpha= 270.2^\circ$, $\beta = 98.0^\circ$ and $\gamma = 0^\circ$. We note that the unperturbed rank 2 crystal-field parameter has a magnitude of $B^2_0 = 578$~cm$^{-1}$, such that the observed change in the rank 2 crystal-field potential consisted of around $\sim2$\% for both of the ions. This amounts to a frequency shift of $3.5$~GHz for Ion 1 and $3.9$~GHz for Ion 2 for the $^4$I$_{15/2}$Z$_1$ to $^4$I$_{13/2}$Y$_1$ transition when compared to the crystal-field model not including any perturbation, which is somewhat larger than, but comparable to, the observed inhomogeneous linewidth for single ions coupled to the nanophotonic device. Therefore, the observed $g$-value anisotropy is consistent with a small amount of strain near the ion, potentially due to proximity to the surface or a charge compensation mechanism. 

In order to perform the initial crystal-field Hamiltonian fit we independently measured the $^4$I$_{15/2}$Z$_1$ and $^4$I$_{13/2}$Y$_1$ ensemble magnetic moments using optical spectroscopy by utilizing the high density implanted sample. From this we obtain $g_{\perp} = 8.6$ and $g_{\parallel} = 1.4$ for the ground state, and $g_{\perp} = 7.6$ and $g_{\parallel} = 1.3$ for the excited state.
The fitted ground state $g$-values deviate from the literature values measured by electron-spin resonance \cite{Bernal1971}; however, the deviation is within the uncertainty of our vector magnet calibration due to magnetic field gradients within the sample space. We note that the above conclusion about the mechanism for the anisotropy of single ion $g$-values does not depend on the precise  magnetic moments assumed for bulk \ercawo. Additionally, in the above fit we have constrained that $g_x = g_y$, and our data was also consistent with a fit for which $g_x \neq g_y$ at the 5\% level. This suggests that some of the observed distortion of the magnetic moment tensor may also be present in an ensemble average.

\section{Dependence of $T_1$ on magnetic field}
\begin{figure*}[ht]
	\centering
    \includegraphics[]{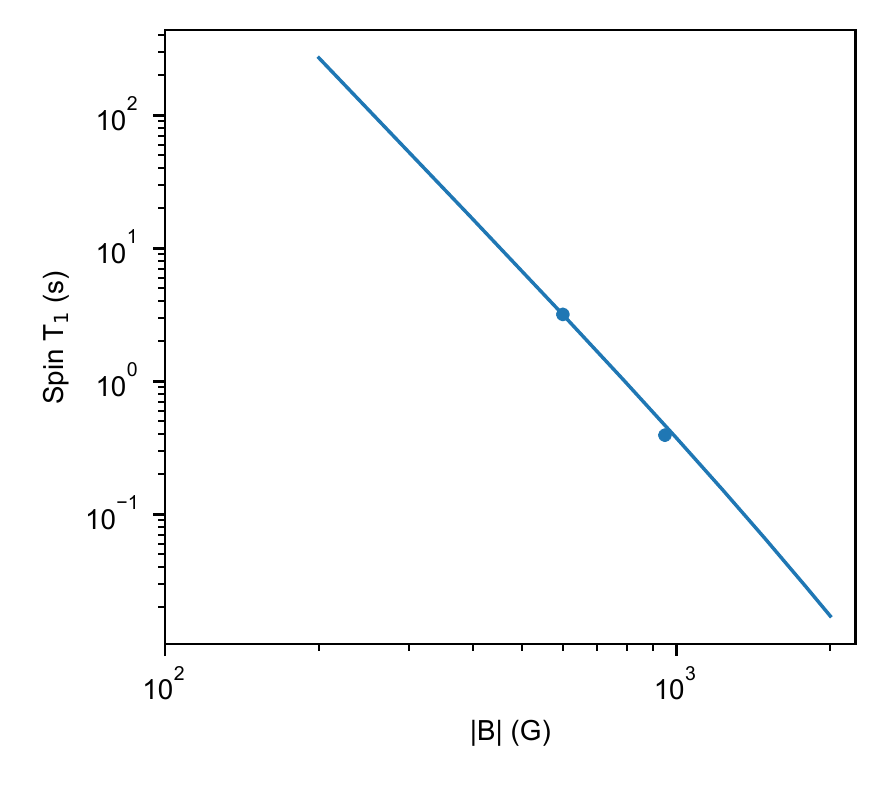}
    \caption{\textbf{Spin lifetime as a function of magnetic field strength}. Solid line is a fit to Eq.\eqref{eq:T1} as is predicted for the spin-lattice relaxation time.}
    \label{fig:T1scaling}
\end{figure*}

To understand the observed spin lifetime, we performed a lifetime measurement at a second magnetic field strength, $|B|=950$~G, obtaining $T_{1}=0.393$~s. At low temperatures, Raman and Orbach processes are slow and the spin relaxation rate is dominated by the direct process. This has the following functional form with respect to an applied magnetic field \cite{Abragam1970}:
\begin{equation}
    T_1^{-1} = A_d ~{\bigg(\frac{g \mu_B B}{h}\bigg)}^5 \coth\bigg(\frac{ g \mu_B B}{2k_bT} \bigg).  
    \label{eq:T1}
\end{equation}
Fitting the above equation to the observed spin lifetime data (Fig.~\ref{fig:T1scaling}) yielded a direct process constant $A_d= 6.4(2) \times10^{-6}$ s$^{-1}$ GHz$^{-5}$, with the expected $T_1 \propto 1/B^5$ scaling. The spin $T_1$ has previously been measured for CaWO$_4$ at low temperatures, and the zero-temperature extrapolated lifetime was found to be $4.8$~s at a frequency of $7.881$~GHz~\cite{LeDantec2021}. This corresponds to $A_d = 7.2 \times 10^{-6}$~s$^{-1}$GHz$^{-5}$, approximately consistent with our value.

\section{Spin coherence modeling}
In order to explain the observed spin coherence, we consider the magnetic environment of the \er ion in the CaWO$_4$ host crystal, consisting of the $^{183}$W nuclear spin bath and paramagnetic impurities. While the concentration and dynamics of paramagnetic impurities is not well known, the dynamics of the nuclear spin bath under decoupling sequences can be understood using standard CCE (Cluster Correlation Expansion) techniques \cite{yang2009quantum}. First, we describe the \er spin coupling to the nuclear spin bath and explain features observed in the Hahn experiment at short times. 
Then, we apply our understanding of the nuclear spin bath to find the expected decoherence rates for the Hahn and Ramsey experiments and observe that it cannot explain the observed rates in either. This leads us to conjecture the existence of an appreciable concentration of paramagnetic impurities, which we discuss in the next section.

To form the nuclear spin bath, we generate random configurations of nuclear spins by allowing each W atom to be an $^{183}$W isotope ($I = 1/2$) with probability 14.3\%. Under the secular approximation for the electron spin, this bath can be described by the following Hamiltonian in the rotating frame of the \er spin:
\begin{equation}
    H = 2S_z\sum_i(A^{(i)}_{||} I^i_z + A^{(i)}_{\perp}I^i_x) + \sum_i\omega_{L,W} I^i_z + \sum_{i,j} H_{nn}^{(i,j)},
\end{equation}
where $I^i_{z/x}$ are the nuclear spin operators of the i$^\text{th}$ spin, $A^{(i)}_{||}$ and $A^{(i)}_{\perp}$ are the parallel and perpendicular hyperfine interaction terms, $\omega_{L,W}$ is the Larmor frequency of the W nuclear spin (107.7~kHz at $|B| = 600$~G) and $H_{nn}^{(i,j)}$ is the dipolar interaction Hamiltonian between W nuclear spins.

This implies that there can be considerable variation in ESEEM (Electron Spin Echo Envelope Modulation) features observed for an \er ion, depending on whether a nearby W nuclear spin is present. We observe such features as dips in coherence in the Hahn experiment (Fig.~4d). In contrast to decay envelopes, these sharp features occur at particular pulse spacings and indicate coherent coupling to a $W$ nuclear spin occupying one of the nearest sites.
Since we are working under the assumption that there is only one nearby W nuclear spin, we do not allow another nuclear spin within the first ten nearest W sites when generating the random nuclear spin baths. We find hyperfine parameters of the strongly coupled W nuclear spin by minimizing over the following cost function:
\begin{equation}
    C(A_{||},A_{\perp}) = \sum_k \sum_j (S_{\text{sim},k}(\tau_j) - S_{\text{exp}}(\tau_j) )^2;\quad 
    S_{\text{sim},k}(\tau_j) = e^{-(2\tau_j/T_2)^n} \cdot S_{\text{bath},k}(\tau_j) \cdot S(\tau_j,A_{||},A_{\perp}).
\end{equation}
Here, $\tau_j$ are the time units that were sampled in the Hahn experiment, $S_{\text{sim},k}(\tau_j)$ is the simulated signal for the k$^\text{th}$ bath and $S_{\text{exp}}(\tau_j)$ is the experimentally measured contrast for the experiment. $S_{\text{sim},k}(\tau_j)$ is calculated by taking a product of three factors: a stretched exponential decay, with decay constants $T_2 = 44$~$\mu$s and $n=1.4$ used in Fig.~4d, the CCE simulation for the nuclear spin bath, $S_{\text{bath,k}}(\tau_j)$, and simulation for a single W nuclear spin coupled to Er$^{3+}$, $S(\tau_j,A_{||},A_{\perp})$. The form of $S_{\text{sim},k}(\tau_j)$ is justified as a first order CCE expansion, which does not take interactions between constituents of the bath into account. The envelope $e^{-(2\tau_j/T_2)^n}$ is assumed to come from a source independent of the nuclear spin bath.

The minimization yields the hyperfine parameters, $(A_{||},A_{\perp})$ = (25.2, 31.7)~kHz. These values are within range of expected interaction strengths for nearby W nuclear spin. In particular, for the W nuclear spin coordinates given as $\mathbf{r_W} = \pm\mathbf{a}/2+\mathbf{c}/2$, where $\mathbf{a}= a\hat{x}$ and $\mathbf{c}=c\hat{z}$ are CaWO$_4$ lattice vectors, we calculate $(A_{||},A_{\perp})$ = (15.5, 30.5)~kHz at our field orientation. The discrepancy could be due to uncertainty in the field alignment or the \er spin g-tensor, which can lead to rotations as discussed in Sec.~\ref{sec:g-factor}.

\begin{figure*}[ht]
	\centering
    \includegraphics[width= 90 mm]{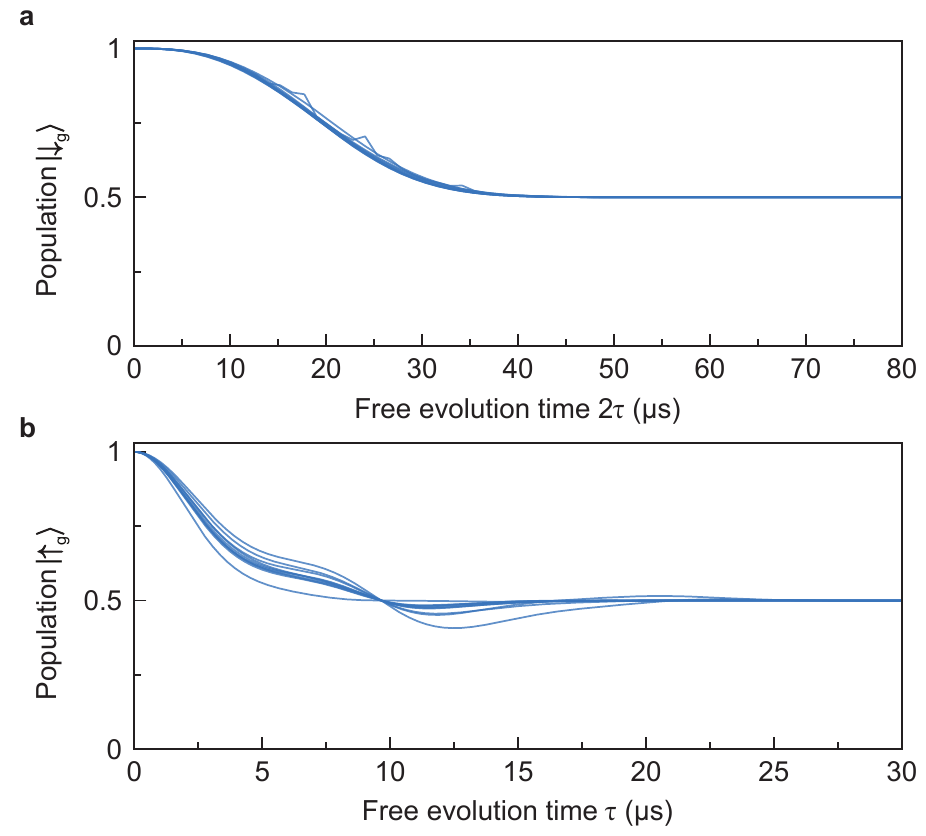}
    \caption{\textbf{W bath limited coherence}.
    \textbf{a} The second order contribution to the CCE simulation for the Hahn experiment for 10 random W-bath configurations, where we assume that the nearest W nuclear spin is located at $\mathbf{r_W}$. Fitting each of the curves to a stretched exponential yields $T_2 = 22.7(4)$~ms with $n=2.7(1)$. We note that this is only the envelope and faster ESEEM features, obtained from the first order CCE simulation, persist as seen in Fig.~4d. This simulation considers W nuclear spins within an 11~nm radius of the \er spin.
    \textbf{b} CCE simulation of Ramsey experiment for the same W bath configurations. Fitting each of the curves to a Gaussian decay yields $T_2^* = 4.0(4)$~$\mu$s. Both simulations are performed at our experimental field configuration.
    }
    \label{fig:wBathCoherence}
\end{figure*}

Finally, ignoring the contribution from the phenomenological decay, we simulate longer time delays to extract the W bath limited coherence times for the Hahn and Ramsey experiments. We perform a second order CCE simulation for the simulation of the Hahn experiment, which takes into account the dipolar coupling between nuclear spins. Noting that ESEEM features will persist as observed in Fig.~4d, we find that the interaction between nuclear spin pairs leads to an envelope which decays in 22.6~ms (Fig.~\ref{fig:wBathCoherence}a). We also perform a Ramsey simulation and show that the expected $T_2^*$ decoherence due to the W-bath is about 4~$\mu$s (Fig.~\ref{fig:wBathCoherence}b). Both of these coherence values are significantly longer than the observed values for $T_2$ and $T_2^*$. We attribute the difference to paramagnetic impurities and explore the expected concentration in the next section.

\section{Estimating concentration of paramagnetic impurities}

In order to estimate the concentration of paramagnetic impurities, we use the Ramsey experiment as a probe of the static magnetic noise experienced by the Er$^{3+}$ ion. As stated in the previous sections, the W bath limited $T_2^*$ (Fig.~\ref{fig:wBathCoherence}b) is significantly longer than the measured $T_2^*$ of 247 ns. This indicates that the measured $T_2^*$ is limited by paramagnetic impurities. Therefore, we can use the measured $T_2^*$ to roughly estimate the concentration of these impurities.

Without loss of generality, we can assume that the interaction between the \er spin and the paramagnetic impurity will be of the Ising form under the secular approximation, forbidding exchanges that do not conserve magnetization. Under these assumptions, we can write down the Hamiltonian concerning the \er spin and the paramagnetic bath in the frame rotating at \er and the impurity frequency for a single impurity species
\begin{equation}
    H = S_z\sum_i J_{I}^{(i)}S^i_z + \sum_{i,j}J_{I}^{(i,j)}S^i_zS^j_z + J_{S,k}^{(i,j)}(S^i_xS^j_x + S^i_yS^j_y) + \sum_i \Delta_iS^i_z.
\end{equation}
Here, $J_{I}^{(i)}$ is the Ising interaction strength between the \er spin and the impurity, while $J_{I}^{(i,j)}$ and $J_{S}^{(i,j)}$ are the Ising and Symmetric interaction strengths between the two impurities and $\Delta_i$ is the disorder in the precession frequency of each impurity. While the bath interaction and disorder terms become important for decoupling sequences, these processes do not contribute on the timescale of the Ramsey experiment, which is dominated by the static noise represented by the first term. Based on this understanding, we can compute the net frequency, in the rotating frame, of the \er spin given the state of the bath. For the purposes of this calculation, we take this bath to consist of $S=1/2$ electrons with $g=2$ and represent its state by the bitstring $k$ of length $N$. We can then write down the Hamiltonian projected by the state $k$ of the bath
\begin{equation}
    H_k = \langle k| S_z\sum_i J_{I}^{(i)}S^i_z |k\rangle = 
    S_z \sum_i J_{I}^{(i)} \frac{(-1)^{k_i}}{2} = \omega_k S_z;\quad
    \omega_k = \sum_i\frac{(-1)^{k_i}J_{I}^{(i)}}{2}.
\end{equation}
Here, $\omega_k$ is the precession frequency of the \er spin given the state $k$ of the bath. We can then calculate the Ramsey coherence as $T_2^* = \pi / \Delta\omega$, where $\Delta\omega^2$ is the variance over the set $\{\omega_k\}$:
\begin{equation}
\begin{split}
    \Delta\omega^2 &= \frac{1}{2^N} \sum_k \omega_k^2 = 
    \frac{1}{2^N} \sum_k \sum_i\Big{(}\frac{(-1)^{k_i}J_{I}^{(i)}}{2}\Big{)}^2 + \frac{1}{2^N}\sum_k\sum_{i<j} (-1)^{k_i+k_j}J_{I}^{(i)}J_{I}^{(j)} \\
    &= \sum_i \Big{(}\frac{J_{I}^{(i)}}{2}\Big{)}^2,
\end{split}
\end{equation}
where we have used the fact that the distribution is centered at zero and the summations can be reordered. We observe that the frequency standard deviation can be interpreted as the norm of the vector of Ising interaction strengths.

\begin{figure*}[ht]
	\centering
    \includegraphics[width= 94 mm]{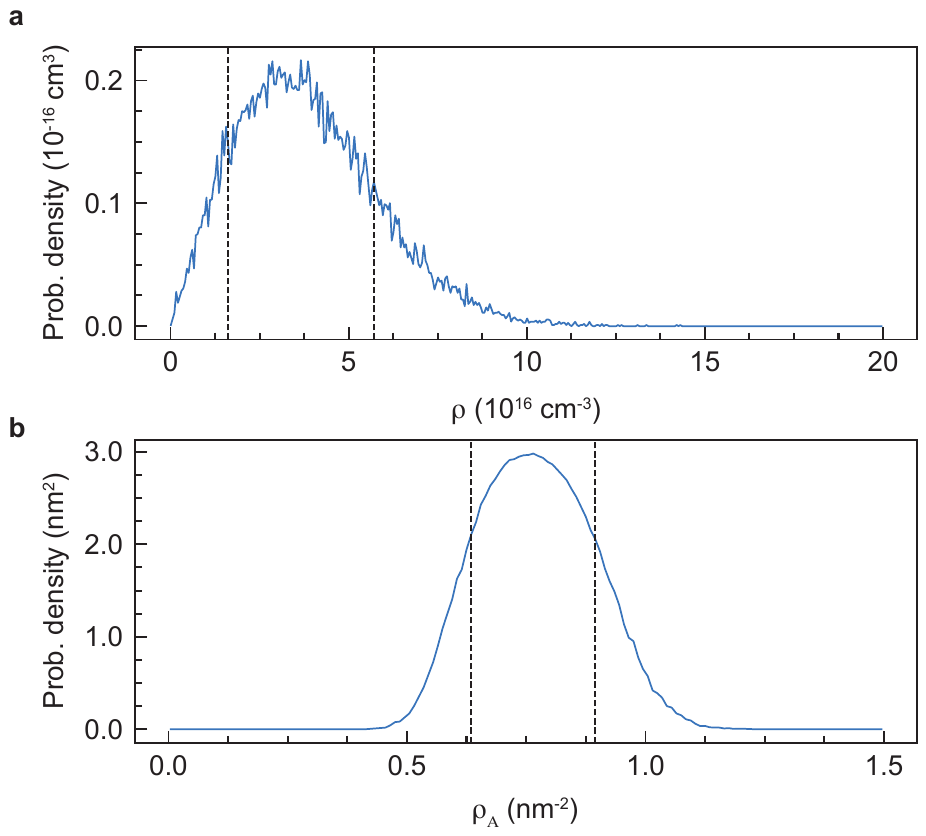}
    \caption{\textbf{Probability density of paramagnetic impurity concentration}.
    \textbf{a} Assuming a 3D uniform distribution of impurities, we estimate that the bath concentration is in the range 1.6$\times10^{16}$ -- 5.7$\times10^{16}$ cm$^{-3}$ with 70\% confidence, with the likeliest concentration at 3.7$\times10^{16}$ cm$^{-3}$.
    \textbf{b} Assuming a 2D distribution on the surface of our crystal, assumed to be located 10 nm away from the \er spin, we estimate an area concentration in the range of 0.5 -- 1.3 nm$^{-2}$ with 70\% confidence, with the likeliest concentration at 0.77 nm$^{-2}$. Dashed lines indicate the confidence ranges for the impurity concentrations.
    }
    \label{fig:para_concentration}
\end{figure*}

In order to estimate the concentration based on this expression, we generate 2000 instances of randomly distributed electron spin baths across a range of concentrations, calculate the resultant $T_2^*$ of each configuration and use Bayes rule to infer a probability density, $P(\rho|T_2^*)$, for the concentration, $\rho$, given our observation of $T_2^*$
\begin{equation}
    P(\rho|T_2^*) = \frac{P(T_2^*|\rho)}{\int_{0}^{\infty} P(T_2^*|\rho^{\prime}) d\rho^{\prime}}.
\end{equation}
Here, $P(T_2^*|\rho^{\prime})$ is the probability to obtain a given $T_2^*$ for the bath concentration $\rho$. In practice, we calculate this expression by counting the number of instances for each concentration that yields a value of $T_2^*$ within 3 standard deviations of the observed value. We perform this estimate for both a 3D geometry of electron spins located in the bulk and a 2D geometry of spins located on the sample surface, estimated to be 10 nm away. We obtain concentrations of $1.6-5.7\times10^{16}$ cm$^{-3}$ and $0.5-1.3$ nm$^{-2}$ for the bulk and surface concentration estimates respectively with 70\% confidence (Fig.~\ref{fig:para_concentration}). As both of these are plausible concentrations, we note that both may be playing a non-negligible role. We are not able to make a calculation of how this concentration of paramagnetic impurities affects the Hahn coherence because it depends on the dynamics of this paramagnetic impurity bath, which is not well known. Doing this calculation would require further information such as the species and spin-lifetime of impurities, as well as the disorder of the bath.

\clearpage

\bibliography{library.bib}